\documentclass[conference, 10pt, twocolumn]{IEEE_Transaction}

\def\BibTeX{{\rm B\kern-.05em{\sc i\kern-.025em b}\kern-.08em T\kern-.1667em\lower.7ex\hbox{E}\kern-.125emX}}
\usepackage{tikz}
\usetikzlibrary{automata,positioning}
\usepackage[T1]{fontenc}
\usepackage{amsmath,amsfonts,amssymb,amsthm,amsbsy}
\usepackage{float}
\usepackage{textcomp}
\usepackage{cite}
\usepackage{bm}
\usepackage{algorithm} 
\usepackage{xcolor}
\usepackage{arydshln}
\usepackage{multicol}
\makeatletter
\newsavebox\myboxA
\newsavebox\myboxB
\newlength\mylenA

\newlength{\punctuationfootlength}
\newcommand{\punctuationfootnote}[2]{#2\settowidth{\punctuationfootlength}%
{#2}\hspace{-0.3\punctuationfootlength}\footnote{#1}}

\newcommand*\xoverline[2][0.75]{%
    \sbox{\myboxA}{$\m@th#2$}%
    \setbox\myboxB\null
    \ht\myboxB=\ht\myboxA%
    \dp\myboxB=\dp\myboxA%
    \wd\myboxB=#1\wd\myboxA
    \sbox\myboxB{$\m@th\overline{\copy\myboxB}$}
    \setlength\mylenA{\the\wd\myboxA}
    \addtolength\mylenA{-\the\wd\myboxB}%
    \ifdim\wd\myboxB<\wd\myboxA%
       \rlap{\hskip 0.5\mylenA\usebox\myboxB}{\usebox\myboxA}%
    \else
        \hskip -0.5\mylenA\rlap{\usebox\myboxA}{\hskip 0.5\mylenA\usebox\myboxB}%
    \fi}
\makeatother
\usepackage[algo2e]{algorithm2e} 
\usepackage[noend]{algpseudocode}
\newcommand\blfootnote[1]{%
	\begingroup
	\renewcommand\thefootnote{}\footnote{#1}%
	\addtocounter{footnote}{-1}%
	\endgroup}
\usepackage{lipsum}
\usepackage{authblk}
\makeatletter
\renewcommand\AB@affilsepx{ \protect\Affilfont} 
\makeatother
\makeatletter
\def\BState{\State\hskip-\ALG@thistlm}
\makeatother
\usepackage{mathtools}

\theoremstyle{remark}

\usepackage{amsmath}

\usepackage[cmintegrals]{newtxmath}

\newtheorem{theorem}{\textbf{Theorem}}
\renewenvironment{proof}{{\textbf{Proof.}}}{}
\begin{document}

	\title{Bit-Interleaved Coded Multiple Beamforming in Millimeter-Wave Massive MIMO Systems }

	\small
	\author{Sadjad Sedighi}
	\author{Ender Ayanoglu}
	\affil{CPCC, Dept of EECS, UC Irvine, Irvine, CA, USA,}
	\normalsize

	\maketitle
	\begin{abstract}
In this paper we carry out the asymptotic diversity analysis for millimeter-wave (mmWave) massive multiple-input multiple-output (MIMO) systems by using bit interleaved coded multiple beamforming (BICMB). First, a single-user mmWave system which employs $M_t$ antenna subarrays at the transmitter and $M_r$ antenna subarrays at the receiver is studied. Each antenna subarray in the transmitter and the receiver has $N_r$ and $N_t$ antennas, respectively. We establish a theorem for the diversity gain when the number of antennas in each remote antenna unit (RAU) goes to infinity. Based on the theorem, the distributed system with BICMB achieves full spatial multiplexing of $L_t=\sum_{i,j}L_{ij}$ and full spatial diversity of $\frac{\left(\sum_{i,j}\beta_{ij}\right)^2}{\sum_{i,j}\beta_{ij}^2L_{ij}^{-1}}$ where $L_{ij}$ is the number of propagation paths and $\beta_{ij}$ is the large scale fading coefficient between the $i$th RAU in the transmitter and the $j$th RAU in the receiver. This result shows that one can increase the diversity gain in the system by increasing the number of RAUs. Simulation results show that, when the perfect channel state information assumption is satisfied, the use of BICMB increases the diversity gain in the system.
	\end{abstract}
	\blfootnote{This work was partially supported by NSF under Grant No. 1547155.}
	
	\section{Introduction}
	After the need for a new generation for mobile systems to handle tens of billions of wireless devices, millimeter-wave (mmWave) communication became an important candidate for Fifth Generation (5G) mobile communication systems \cite{Rappaport2013,Swindle2014,Roh2014}. The mmWave signals face severe penetration loss and path loss compared to signals in current cellular systems (3G or LTE) \cite{Haidar2014}. One of the advantages of the mmWave frequencies is that they enable us to pack more antennas in the same area compared to a lower range of frequencies\punctuationfootnote{Although strictly speaking, mmWave corresponds to the range 30-300 GHz, in practical use, about 5-6 to 100 GHz may be termed mmWave frequencies.}. This leads to highly directional beamforming and large-scale spatial multiplexing in mmWave frequencies. By exploiting these properties, massive MIMO systems can be developed. The principles of beamforming are independent of the carrier frequency, but it is not practical to use fully digital beamforming schemes for massive MIMO systems \cite{Telatar1999,Shi2011,Dahrouj2010,Peel2005}. Power consumption and cost perspectives are the main obstacles due to the high number of radio frequency (RF) chains required for the fully digital beamforming, i.e., one RF chain per antenna element \cite{Doan2004}. To address this problem, hybrid analog-digital processing of the precoder and combiner for mmWave communications systems is being considered \cite{Sohrabi2016,Ayach2012,Ayach2013,Weiheng2016,Singh2014,Zhang2015}. 
	
	Bit-interleaved coded modulation (BICM) was introduced as a way to increase the code diversity \cite{Caire1998,Zehavi1992}.  As stated in \cite{Akay2007}, bit-interleaved coded multiple beamforming (BICMB) has a great impact on the diversity gain performance of a MIMO system. Recently, by comparing the diversity gain in both co-located and distributed systems the authors of \cite{Dian2018} have shown that  increasing the number of RAUs in the distributed system does increase the diversity gain and/or multiplexing gain. In Section II and III, BICMB is analyzed. We show that by using BICM in the system, one can achieve full spatial multiplexing without any loss in the diversity gain. That is, in Section III, we show that BICMB achieves full diversity order of $M_rM_tL$ and full spatial diversity of $M_rM_tL$ in a special case when the number of propagation paths is constant for all paths between RAUs, i.e., $L_{ij}=L$ over a limited scattering mmWave channel. We provide design criteria for the interleaver that guarantee full diversity and full spatial multiplexing.

We would like to reiterate that the asymptotical diversity analysis obtained in this paper is under the idealistic assumption of having perfect channel state information both at the transmitter and the receiver as done in similar works.

	\textit{Notation:}
Boldface upper and lower case letters denote matrices and column vectors, respectively.	The minimum Hamming distance
	of a convolutional code is defined as $d_{\text{free}}$. The symbol $N_s$ denotes
	the total number of symbols transmitted at a time.
	The minimum Euclidean distance between the two constellation points is given by $d_{\text{min}}$. The superscripts $(.)^H, (.)^T, (.)^*,(\bar{.})$ and the symbol $\forall$ denote the Hermitian, transpose, complex conjugate, binary complement, and for all, respectively. $\mathcal{CN} (0, 1)$ denotes a circularly symmetric complex Gaussian random variable with zero mean and unit variance. The expectation operator is denoted by $E\left[.\right]$. Also, $[\mathbf{A}]_{ij}$ gives the $(i, j)$th entry of matrix $\mathbf{A}$. Finally, diag$\{a_1 , a_2 , \dots , a_N \}$ stands for a diagonal matrix with diagonal elements $\{a_1 , a_2 ,\dots , a_N \}$.
	
	\section{System Model}
\begin{figure} 
		\centering
		\includegraphics[width=.48\textwidth]{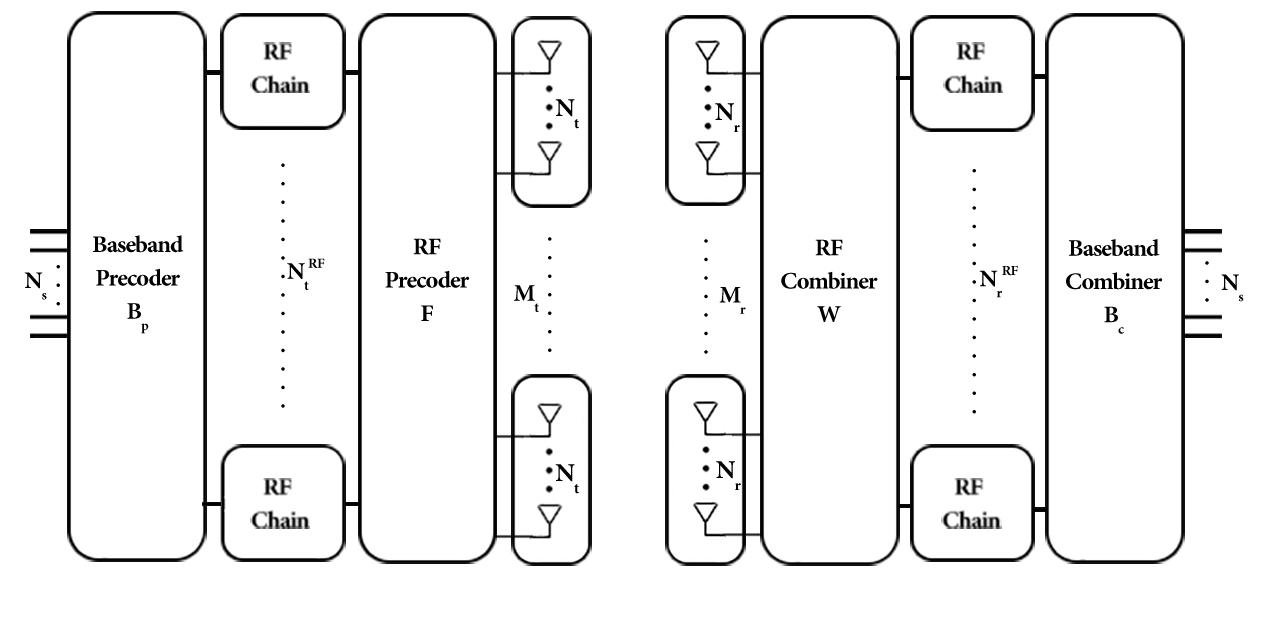}
		\caption{Block diagram of a mmWave massive MIMO system with distributed antenna arrays.}
		\label{SU_FC}
	\end{figure}

	Consider a single-user mmWave massive MIMO system as shown in Fig. \ref{SU_FC}. In this system, the transmitter sends $N_s$ data streams to a receiver. The transmitter is equipped with  $N_t^{RF}$ RF chains and $M_t$ RAUs, where each RAU has $N_t$ antennas, while at the receiver, the number of RF chains and RAUs is given by $N_r^{RF}$ and $M_r$, respectively. Each RAU at the receiver has $N_r$ antennas. When $M_t = M_r = 1$, the system reduces to a conventional co-located MIMO (C-MIMO) system.

	The input to the system is $N_s$ data streams. The vector of data symbols to be transmitted by the transmitter at each time instant, $\mathbf{x} \in \mathbb{C}^{N_s\times1}$, can be expressed as
	\begin{align}
	\mathbf{x}=\left[x_1,...,x_{N_s}\right]^T,
		\end{align} 
		where $E\left[\mathbf{x}\mathbf{x}^H\right]=\mathbf{I}_{N_s}$. The preprocessing at the baseband is applied by means of the matrix $\mathbf{B}_p \in \mathbb{C}^{N_t^{RF} \times N_s}$. The last stage of data preprocessing is performed at RF, when beamforming is applied by means of phase shifters and combiners. A set of $M_tN_t$ phase shifters is applied to the output of each RF chain. As a result of this process, different beams are formed in order to transmit the RF signals. We can model this process with an $M_t N_t \times N_t^{RF} $ complex matrix $\mathbf{F}$. Note that the baseband precoder $\mathbf{B}_p$ modifies both amplitude and phases, while only phase changes can be realized by $\mathbf{F}$ since it is implemented by using analog phase shifters. 
		
We assume a narrowband flat fading channel model and obtain the received signal	as
	\begin{align}
	\mathbf{z}=\mathbf{HF}\mathbf{B}_p\mathbf{x}+\mathbf{n},
	\end{align}
	where $\mathbf{H}$ is an $M_rN_r\times M_tN_t$ channel matrix with complex-valued entries and $\mathbf{n}$ is an $M_rN_r\times1$ vector consisting of i.i.d. $\mathcal{CN}(0,N_0)$ noise samples, where $N_0=\frac{N_t}{SNR}$. The processed signal is given by
	\begin{align}\label{rec_sig}
	\mathbf{y}=\mathbf{B}_c^H\mathbf{W}^H\mathbf{HF}\mathbf{B}_p\mathbf{x}+\mathbf{B}_c^H\mathbf{W}^H\mathbf{n},
	\end{align} 
	where $\mathbf{W}$ is the $M_rN_r\times N_r^{RF}$ RF combining matrix, and $\mathbf{B}_c$ is the $N_r^{(rf)}\times N_s$ baseband combining matrix.
	
	The channel matrix $\mathbf{H}$ can also be written as
	\begin{align}
	\mathbf{H}=
	\begin{bmatrix}
	\sqrt{\beta_{11}}\mathbf{H}_{11}& \dots& \sqrt{\beta_{1M_t}}\mathbf{H}_{1M_t} \\
	\vdots & \ddots & \vdots  \\
	\sqrt{\beta_{M_r1}}\mathbf{H}_{M_r1} & \dots & \sqrt{\beta_{M_rM_t}}\mathbf{H}_{M_rM_t}
	\end{bmatrix},
	\end{align}
	where $\beta_{ij}$, a real-valued nonnegative number, represents the large-scale fading effect between the $i$th RAU at the receiver and $j$th RAU at the transmitter. The normalized subchannel matrix $\mathbf{H}_{ij}$ is the MIMO channel between the $i$th RAU at the receiver and the $j$th RAU at the transmitter.
	
	Analytical channel models such as Rayleigh fading are not suitable for mmWave channel modeling. The reason for this is the fact that the scattering levels represented by these models are too rich for mmWave channels\cite{Ayach2012}. In this paper, the model is based on the Saleh-Valenzuela model that is often used in mmWave channel modeling \cite{HXu2002} and standardization \cite{Standard}. For simplicity, each scattering cluster is assumed to contribute a single propagation path. The subchannel matrix $\mathbf{H}_{ij}$ is given by
	\begin{align}\label{Hij}
	\mathbf{H}_{ij}=\sqrt{\frac{N_tN_r}{L_{ij}}} \sum_{l=1}^{L_{ij}}\alpha_{ij}^l\mathbf{a}_r(\theta_{ij}^{l}) \mathbf{a}_t^H(\phi_{ij}^{l}),
	\end{align}
	where $L_{ij}$ is the number of propagation paths and $\alpha_{ij}^{l}$ is the complex-gain of the $l$th ray which follows $\mathcal{CN}(0,1)$, the vectors $\mathbf{a}_r(\theta_{ij}^{l})$ and $ \mathbf{a}_t(\phi_{ij}^{l})$ are the normalized receive/transmit array response and $\theta_{ij}^{l} $ and $\phi_{ij}^{l}$ are its random azimuth angles of arrival and departure respectively. The elevation dimension is ignored.  
	
	The uniform linear array (ULA) is employed by the transmitter and receiver in our study. For an $N$-element ULA, the array response vector can be given by
	\begin{align}
	\mathbf{a}_{ULA}(\phi)=\frac{1}{\sqrt{N}}\left[1, e^{j\frac{2\pi}{\lambda}dsin(\phi)},\dots, e^{j(N-1)\frac{2\pi}{\lambda}dsin(\phi)}\right]^T,
	\end{align}
	where $\lambda$ is the wavelength of the carrier, and $d$ is the distance between neighboring antenna elements.

	We leverage both BICM and multiple beamforming to form BICMB.  An interleaver is used to interleave the output bits of a binary convolutional encoder. Then the output of the interleaver is mapped over a signal set $\chi \subset \mathbb{C}$ of size $|\chi|=2^m$ with a binary labeling map $\mu:\{0,1\}^m\rightarrow \chi$. The interleaver design has two criteria\cite{Akay2007}:
	\begin{enumerate}
		\item Consecutive coded bits are mapped over different symbols, 
			\item each subchannel should be utilized at least once within $d_{\text{free}}$ distinct bits among different codewords by using proper code and interleaver. 
	\end{enumerate}
	Note that the free distance $d_{\text{free}}$ of the convolutional encoder should satisfy $d_{\text{free}}\geq N_s$.

For mapping the bits onto symbols, Gray encoding is used. Also, we are using a Viterbi decoder at the receiver. The interleaver $\pi$ is used to interleave the code sequence $\underline{c}$. Then the output of the interleaver is mapped onto the signal sequence $\underline{x} \in \chi$. 

The only beamforming constraint here is a total power constraint, because one can control both the amplitude and the phase of a signal. As we know the total power constraint leads to a simple solution based on  SVD \cite{Ayach2012}
	\begin{align}
	\mathbf{H}=\mathbf{U\Lambda V}^H=\left[\mathbf{u}_1 \mathbf{u}_2 \dots \mathbf{u}_{M_rN_r}\right]^H \mathbf{H} \left[\mathbf{v}_1 \mathbf{v}_2 \dots \mathbf{v}_{M_tN_t}\right],
	\end{align}
	where $\mathbf{U}$ and $\mathbf{V}$ are $M_rN_r\times M_rN_r$ and $M_tN_t\times M_tN_t$ unitary matrices, respectively, and $\mathbf{\Lambda}$ is an $ M_rN_r \times M_tN_t$ diagonal matrix with singular values of $\mathbf{H}$, $\lambda_i \in \mathbb{R}$, on the main diagonal with decreasing order.
By exploiting the optimal precoder and combiner, the system input-output relation in (\ref{rec_sig}) at the $k$th time instant can be written as
	\begin{align}
	\mathbf{y}_k=\left[\mathbf{u}_1 \mathbf{u}_2 \dots \mathbf{u}_{N_s}\right]^H \mathbf{H} \left[\mathbf{v}_1 \mathbf{v}_2 \dots \mathbf{v}_{N_s}\right] \mathbf{x}+ \left[\mathbf{u}_1 \mathbf{u}_2 \dots \mathbf{u}_{N_s}\right]^H\mathbf{n}_k,
	\end{align}
	\begin{align}\label{def_rec}
	y_{k,s}=\lambda_sx_{k,s} +n_{k,s}, \text{   \quad for } s=1,2,\dots,N_s.
	\end{align}

	\section{Diversity Gain and PEP Analysis}
	In this section, we show that by using the BICMB analysis for calculating BER, the diversity gain becomes independent of the number of data streams.
	{\color{black}
		\begin{theorem}
			Suppose that $N_r\rightarrow \infty$ and $N_t \rightarrow \infty$. Then the bit interleaved coded distributed massive MIMO system can achieve a diversity gain of
		\begin{align}\label{su_dg}
		D_g=\frac{\left(\sum_{i,j}\beta_{ij}\right)^2}{\sum_{i,j}\beta_{ij}^2L_{ij}^{-1}}
		\end{align} 
			$\text{ for } i=1,\dots,M_r \text{ and } j=1,\dots,M_t$.
	\end{theorem}}

	\begin{proof}
			We model the BICMB bit interleaver as $\pi: k' \rightarrow (k,s,i)$, where $k'$ represents the original ordering of the coded bits $c_{k'}$, $k$ represents the time ordering of the signals $x_{k,s}$ and $i$ denotes the position of the bit $c_{k'}$ on symbol $x_{k,s}$.
			
		We define $\chi_b^i$ as the subset of all signals $x\in\chi$. Note that the label has the value $b\in\{0,1\}$ in position $i$.
		
		Then, the ML bit metrics are given by using (\ref{def_rec}), \cite{Akay2007,Caire1998,Zehavi1992}
		\begin{align}\label{ML_bit}
		\gamma^i(y_{k,s},c_{k'})=\min_{x \in \chi^i_{c_{k'}}} \left|y_{k,s}-\lambda_sx\right|^2.
		\end{align}
		
		The receiver uses an ML decoder to make decisions based on 
		\begin{align}\label{ML_dec}
		\mathbf{\underline{\hat{c}}}=\text{arg}\min_{\mathbf{\underline{c}}\in\mathcal{C}}\sum_{k'}\gamma^i(y_{k,s},c_{k'}).
		\end{align}
		
		Assume that the code sequence $\underline{c}$ is transmitted and $\underline{\hat{c}}$ is detected. Then by using (\ref{ML_bit}) and (\ref{ML_dec}), the pairwise error probability (PEP) of $\underline{c}$ and $\underline{\hat{c}}$ given channel state information (CSI) can be written as \cite{Akay2007}
		{\color{black}
			\begin{align}
			P&(c\rightarrow \hat{c}|\mathbf{H})\nonumber \\
			=&P\left(\sum_{k'} \min_{x \in \chi^i_{c_{k'}}} |y_{k,s}-\lambda_sx|^2 \geq \sum_{k'} \min_{x \in \chi^i_{\hat{c}_{k'}}} |y_{k,s}-\lambda_sx|^2\right),
			\end{align}}
		where $s \in \{1,2,\dots,N_s\}$.
	
		Note that in a convolutional code, the Hamming distance between $\mathbf{c}$ and $\hat{\mathbf{c}}$, $d(\mathbf{c-\hat{c}})$ is at least $d_{\text{free}}$. In this work we assume for PEP analysis $d(\mathbf{c-\hat{c}})=d_{\text{free}}$. 
		
		For the $d_{\text{free}}$ bits, let us denote
		\begin{align}
		\tilde{x}_{k,s}=\text{arg}\min_{x \in \chi^i_{{c}_{k'}}}\left|y_{k,s}-\lambda_sx\right|^2\\
		\hat{x}_{k,s}=\text{arg}\min_{x \in \chi^i_{\bar{c}_{k'}}}\left|y_{k,s}-\lambda_sx\right|^2
		\end{align}
		
		By using the trellis structure of the convolutional codes \cite{Akay2007}, one can write
		\begin{align}
		P(\mathbf{c}\rightarrow\mathbf{\hat{c}}|\mathbf{H})\leq Q\left(\sqrt{\frac{d_{\text{min}}^2\sum_{s=1}^{N_s}\alpha_s\lambda_s^2}{2N_0}}\right)
		\end{align}
		where $\alpha_s$ is a parameter that indicates how many times subchannel $s$ is used within the $d_{\text{free}}$ bits under consideration, and  $\sum_{s=1}^{N_s}\alpha_s=d_{\text{free}}$.
		There is an upper bound for the $Q$ function $Q(x)\leq\frac{1}{2}e^{-\frac{x}{2}}$, which can be used to upper bound the PEP as
		\begin{align}\label{PEP_exp}
		P(\mathbf{c}\rightarrow\mathbf{\hat{c}})= \mathit{E}\left[P(\mathbf{c}\rightarrow\mathbf{\hat{c}}|\mathbf{H})\right]
		\leq \mathit{E}\left[\frac{1}{2}\text{exp}\left(\frac{-d_{\text{min}}^2\sum_{s=1}^{N_s}\alpha_s\lambda_s^2}{4N_0}\right)\right].
		\end{align}
		
		Let us denote $\alpha_{\text{min}} =\text{min }\{\alpha_s :s=1,2,\dots,N_s\}$. Then
		\begin{align}\label{sing_values}
		\frac{\left(\sum_{s=1}^{N_s}\alpha_s\lambda_s^2\right)}{N_s}\geq \frac{\left(\alpha_{\text{min}}\sum_{s=1}^{N_s}\lambda_s^2\right)}{N_s} \geq \frac{\left(\alpha_{\text{min}}\sum_{s=1}^{L_t}\lambda_s^2\right)}{L_t}
		\end{align}
		There are only $L_t$ non-zero singular values \cite{Dian2018}.
				
		Let us define
		\begin{align} \label{theta}
		\Theta \triangleq \sum_{s=1}^{N_s}\lambda_s^2=||\mathbf{H}||_F^2=\sum_{i=1}^{M_r}\sum_{j=1}^{M_t}\beta_{ij}||\mathbf{H}_{ij}||_F^2.
		\end{align}
		
		Theorem 3 in \cite{Ayach2012} implies that the singular values of $\mathbf{H}_{ij}$ converge to $\sqrt{\frac{N_rN_t} {L_{ij}}}\left|\alpha_{l}^{ij}\right|$ in a descending order. By using the singular values of $\mathbf{H}_{ij}$, (\ref{theta}) can be rewritten as
		
		\begin{align}\label{Psi_def}
		\Theta=\sum_{i=1}^{M_r}\sum_{j=1}^{M_t}\beta_{ij}||\mathbf{H}_{ij}||_F^2= N_rN_t\sum_{i=1}^{M_r}\sum_{j=1}^{M_t}\underbrace{\frac{\beta_{ij}}{L_{ij}}\sum_{l=1}^{L_{ij}}\left|\alpha_{ij}^l\right|^2}_{\Psi_{ij}}.
		\end{align}
		
		It can be seen easily that $\Psi_{ij}$ in (\ref{Psi_def}) has Gamma distribution with shape $k_{ij}=L_{ij}$ and scale $\theta_{ij}=2\beta_{ij}L_{ij}^{-1}$, i.e., $\Psi_{ij} \sim \mathcal{G}(L_{ij},2\beta_{ij}L_{ij}^{-1})$\cite{Hogg1978}. 
		One can use the Welch-Satterthwaite equation to calculate an approximation to the degrees of freedom of $\Theta$ (i.e., shape and scale of the Gamma distribution) which is a linear combination of the independent random variables $\Psi_{ij}$ \cite{Satterth1946}
		
		\begin{align}\label{shape}
		k&=\frac{\left(\sum_{i,j}\theta_{ij}k_{ij}\right)^2}{\sum_{i,j}\theta_{ij}^2k_{ij}}=\frac{\left(\sum_{i,j}\beta_{ij}\right)^2}{\sum_{i,j}\beta_{ij}^2L_{ij}^{-1}},\\\label{scale}
		\theta&=\frac{\sum_{i,j}\theta_{ij}^2k_{ij}}{\sum_{i,j}\theta_{ij}k_{ij}}=\frac{\sum_{i,j}\beta_{ij}^2L_{ij}^{-1}}{\sum_{i,j}\beta_{ij}}.
		\end{align}
		
		Using (\ref{PEP_exp}), (\ref{sing_values}), and (\ref{theta}), the PEP is upper bounded by
		\begin{align}\label{PEP_exp2}
		P(\mathbf{c}\rightarrow\mathbf{\hat{c}})\leq\frac{1}{2} \mathit{E}\left[\text{exp}\left(\frac{-d_{\text{min}}^2\alpha_{\text{min}}N_s}{4N_0L_t}\Theta\right)\right],
		\end{align}
		which is the definition of the moment generating function (MGF)\cite{Bulmer1965} for the random variable $\Theta$. By using the definition, (\ref{PEP_exp2}) can be written as
		\begin{align}\nonumber
		P(c\rightarrow\hat{c})=&g(d, \alpha_{\text{min}}, \chi)\\ 
		\leq&\frac{1}{2} \left(1+ \theta \frac{d_{\text{min}}^2\alpha_{\text{min}}N_sN_t}{4L_t}SNR\right)^{-k}\label{PEP_MGF1}\\
		\approx&\frac{1}{2}\left(\theta\frac{d_{\text{min}}^2\alpha_{\text{min}}N_sN_t}{4L_t}SNR\right)^{-k} \label{PEP_MGF2}
		\end{align}
		for high SNR.  The function $g(d, \alpha_{\text{min}}, \chi)$ denotes the PEP of two codewords with $d(\underline{c}-\underline{\hat{c}}) = d$, with $\alpha_{\text{min}}$ corresponding to $\underline{c}$ and $\underline{\hat{c}}$, and with constellation $\chi$. 
		In (\ref{PEP_MGF1})  $\theta$ and $k$ are defined as (\ref{shape}) and (\ref{scale}). 
		
		In BICMB, $P_b$ can be calculated as \cite{Akay2007}
		\begin{align}\label{BICMB_Pb}
		P_b \leq \frac{1}{k_c}\sum_{d=d_{\text{free}}}^{\infty}\sum_{i=1}^{W_I(d)}g(d,\alpha_{min}(d,i),\chi),
		\end{align}
		where $W_I(d)$ denotes the total input weight of error events at Hamming distance $d$.	Following (\ref{PEP_MGF2}) and (\ref{BICMB_Pb})
		\begin{align} \label{SU_bound}
		P_b \leq\frac{1}{k_c} \sum_{d=d_{\text{min}}}^{\infty}\sum_{i=1}^{W_I(d)} \frac{1}{2}\left(\theta\frac{d_{\text{free}}^2\alpha_{\text{min}}N_sN_t}{4L_t}SNR\right)^{-k}.
		\end{align}
		
		The SNR component has a power of $-k$ for all summations. Hence, BICMB achieves full diversity order of
		\begin{align}\label{su_dg2}
		D_g=k=\frac{\left(\sum_{i,j}\beta_{ij}\right)^2}{\sum_{i,j}\beta_{ij}^2L_{ij}^{-1}}
		\end{align} 
	which is independent of the number of spatial streams transmitted.
	\end{proof}

	\newtheorem{remarks}{\textbf{Remark}}
	\begin{remarks}
		Under the case where $N_t$ and $N_r$ are large enough and assuming that $L_{ij}=L$ and $\beta_{ij}=\beta$ for any $i$ and $j$, it can be seen easily that the distributed massive MIMO system can achieve a diversity gain
		\begin{align}
		G_d=L_t=M_rM_tL.
		\end{align}
		
	\end{remarks}

	\begin{remarks}
		Theorem 1 implies that the diversity gain is independent of the number of data streams, i.e., the transmitter can send the maximum number of data streams $N_s\leq L_t$, and still get the same diversity gain. This will be illustrated in Section IV.
	\end{remarks}
	\section{Simulation Results}
\begin{figure}[ht]
	\centering
	\includegraphics[width=.45\textwidth]{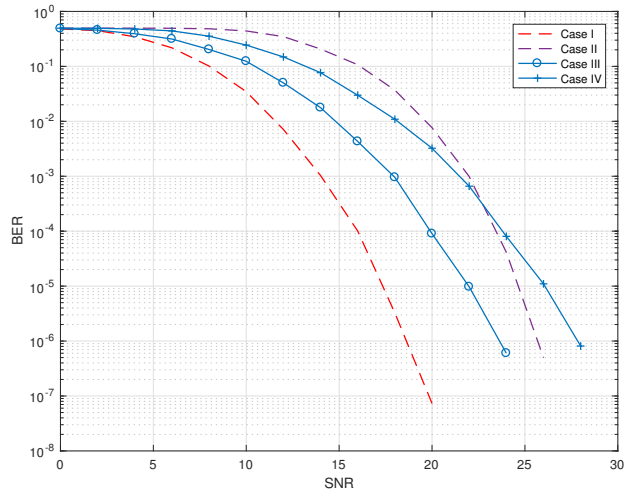}
	\caption{BER with respect to $\text{SNR}$ with $N_t=100$, $N_r=50$ and $N_s=1$.}
	\label{diff_g}
\end{figure}

	In the simulations, the industry standard 64-state 1/2-rate (133,171) $d_{\text{free}}=10$ convolutional code is used. For BICMB, we separate the coded bits into different substreams of data and a random interleaver is being used to interleave the bits in each substream. We assume that the number of RF chains in the receiver and transmitter are twice the number of data streams \cite{Sohrabi2016} (i.e., $N_t^{\text{RF}}=N_r^{\text{RF}}=2N_s$) and each scale fading coefficient $\beta_{ij}$ equals $\beta = −20$ dB (except for Fig. \ref{diff_g}). For the sake of simplicity, only ULA array configuration with $d=0.5$ is considered at RAUs and BPSK modulations is employed for each data stream.
	
\begin{figure}[ht]
	\centering
	\includegraphics[width=.45\textwidth]{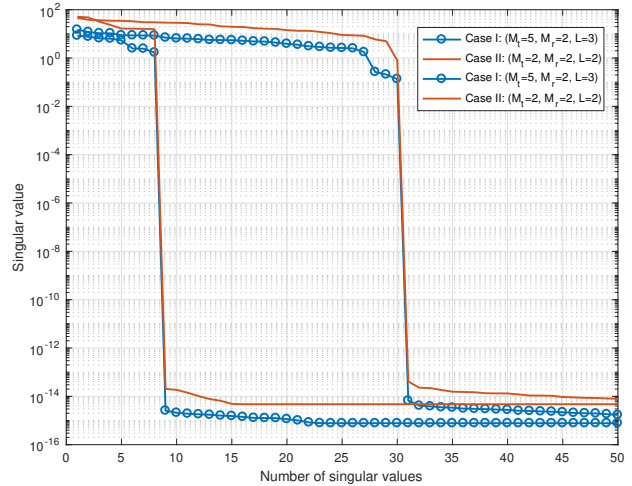}
	\caption{Singular values of the sparse mmWave channel with $N_t=100$ and $N_r=50$.}
	\label{singluar}
\end{figure}

	Two different cases are simulated in Fig. \ref{singluar}. In Case I, $N_t=2N_r=100$, while in Case II, $N_t=2N_r=400$. Fig. \ref{singluar} shows that the number of singular values of the mmWave channel is limited, i.e., there are only limited subchannels which can be used to transmit the data.  The number of available subchannels $L_t=\sum_{i=1}^{M_r}\sum_{j=1}^{M_t}L_{ij}$ which is the rank of the channel $\mathbf{H}$ and is independent of the number of antennas in RAUs in both transmitter and receiver side.

\begin{figure}[ht]
	\centering
	\includegraphics[width=.45\textwidth]{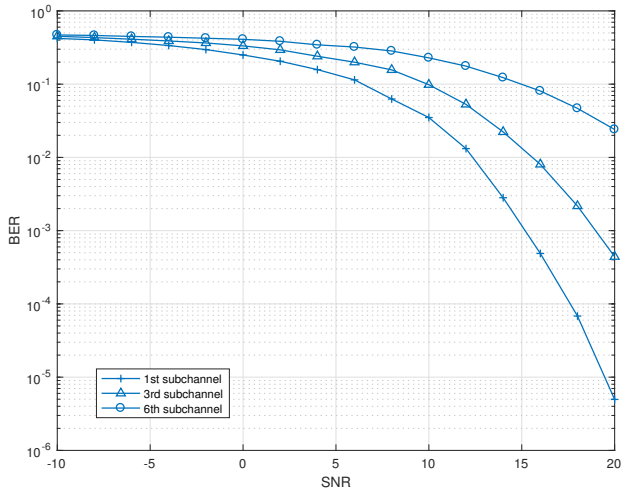}
	\caption{BER with respect to $\text{SNR}$ with $N_t=100$, $N_r=50$, $M_t=2$, $M_r=2$  and $L=2$ for $N_s=6$.}
	\label{no_intlv}
\end{figure}

Fig. \ref{no_intlv} illustrates the importance of the interleaver design. A random interleaver is used such that consecutive coded bits are transmitted over the same subchannel. Consequently, an error on the trellis occurs over paths that are spanned by the worst channel and the diversity order of coded multiple beamforming approaches to that of uncoded multiple beamforming with uniform power allocation. In other words, the BER performance decreases when the interleaving design criteria are not met.  

\begin{figure}[ht]
		\centering
		\includegraphics[width=.45\textwidth]{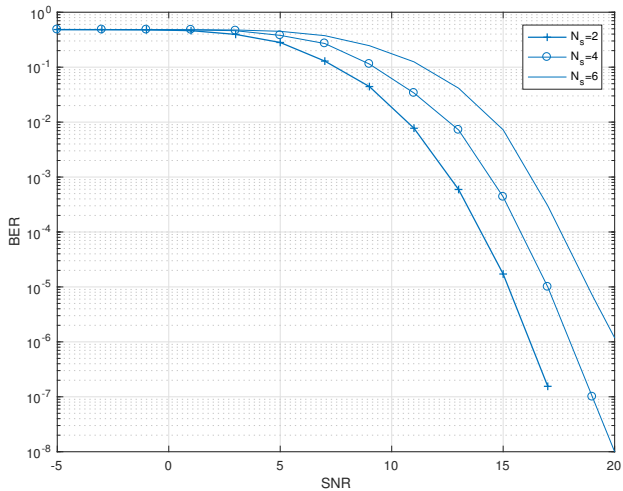}
		\caption{BER with respect to $\text{SNR}$ with $N_t=100$, $N_r=50$, $M_t=3$, $M_r=1$  and $L=2$ for different values of $N_s$.}
		\label{su_dg8}
	\end{figure}

	On the other hand, as we expect from (\ref{su_dg2}), changing the number of streams $N_s$ should not change the diversity gain, i.e., the slope of the BER curve in high SNR. As it can be seen from Fig. \ref{su_dg8}, the slope does not change by changing the number of data streams. Hence, one can get the same diversity gain by using the maximum number of data streams available ($L_t$). 
	
\begin{figure}[ht]
	\centering
	\includegraphics[width=.45\textwidth]{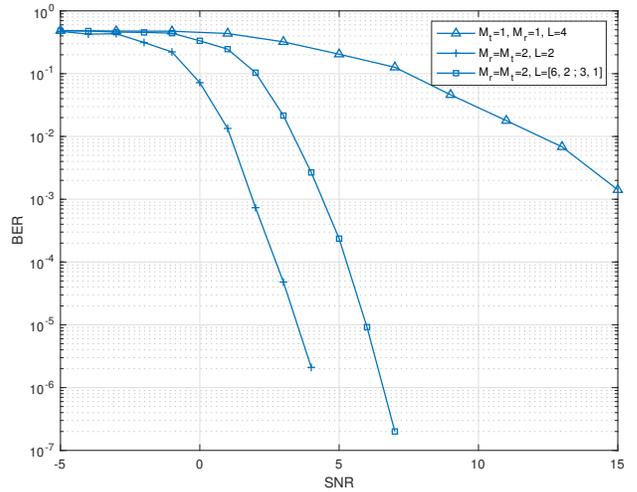}
	\caption{BER with respect to $\text{SNR}$ with $N_t=100$, $N_r=50$ and $N_s=3$.}
	\label{co}
\end{figure}

Fig. \ref{co} illustrates the results for BICMB for both co-located and distributed mmWave massive MIMO systems. The diversity gain for the distributed system outperforms the co-located system, even though the channel in the co-located system has richer scattering (the number of propagation paths in the co-located system is twice as the distributed system). Also, as it can be seen from the figure, the curves for the distributed systems are parallel to each other, especially for the high-SNR region, which can be confirmed by (\ref{su_dg2}). 

Fig. \ref{diff_g} illustrates the effect of the large scale fading coefficient on the diversity gain. Despite other simulations, we consider inhomogeneous large scale fading coefficients. In these simulations, information bits
are mapped onto 16 quadrature amplitude modulation (QAM)
symbols in each subchannel. When $M_r=M_t=2$, $L_{ij}=L=2$, $N_t=2N_r=100$ and $N_s=1$ three different cases are simulated. Let $\mathcal{\mathbf{B}}=\left[\beta_{ij}\right]$ where $\beta_{ij}$ expressed in dB, as the large scale fading coefficient matrix. We used the following $\mathbf{B}$ in the simulations:
\begin{align*}
    &\mathbf{B}_1=\begin{bmatrix}
    -20   & -20 \\
    -20  & -20 
\end{bmatrix},
    \mathbf{B}_2=\begin{bmatrix}
    -25   & -25 \\
    -25  & -25 
\end{bmatrix},\\
    &\mathbf{B}_3=\begin{bmatrix}
    -20   & -35 \\
    -35  & -20
\end{bmatrix},
    \mathbf{B}_4= -20 .
\end{align*}

As it can be seen from Fig. \ref{diff_g}, when the system is homogeneous, the diversity gain remains the same. Case I and Case II, have the same slope in high SNR, which is expected. In Case III, when the system is inhomogeneous, the diversity gain decreases. By using (\ref{su_dg2}), one can easily see that Case III has approximately the same diversity gain as a system with $M_r=M_t=1$ and $L=4$, i.e., $D_G=4$, which is depicted in Case IV.
	\section{Conclusion}
	In this paper we analyzed BICMB in mmWave massive MIMO systems. BICMB achieves full spatial diversity of $\frac{\left(\sum_{i,j}\beta_{ij}\right)^2}{\sum_{i,j}\beta_{ij}^2L_{ij}^{-1}}$ over $M_t$ RAU transmitters and $M_r$ RAU receivers. This means, by increasing the number of RAUs in the distributed system with BICMB, one can increase the diversity gain and multiplexing gain. As it can be seen from the diversity gain formula for the single-user system, the value of diversity gain is independent of the number of antennas in each RAU for both transmitter and receiver. A special case of the diversity gain where $L_{ij}=L$ and $\beta_{ij}=\beta$ would be $M_rM_tL$ which is similar to the diversity gain of a convential MIMO system.
	\bibliographystyle{IEEEtran}
	\bibliography{icc2019}
\end{document}